\documentclass[journal]{IEEEtran}
\usepackage{cite}
\usepackage{amsmath,amssymb,amsfonts}

\usepackage{algorithm} 
\usepackage{subfigure}
\usepackage{textcomp}
\usepackage{bm}
\usepackage{amsmath,amsfonts}
\usepackage{algorithmic}
\usepackage{array}
\usepackage[caption=false,font=normalsize,labelfont=sf,textfont=sf]{subfig}
\usepackage{textcomp}
\usepackage{stfloats}
\usepackage{url}
\usepackage{verbatim}
\usepackage{graphicx}
\usepackage{xcolor}
\usepackage{subcaption}

\def\BibTeX{{\rm B\kern-.05em{\sc i\kern-.025em b}\kern-.08em
    T\kern-.1667em\lower.7ex\hbox{E}\kern-.125emX}}
\usepackage{balance}
\begin{document}
\title{When Digital Twin Meets 6G: Concepts, Obstacles, and Research Prospects}

\author{\author{Wenshuai Liu, Yaru Fu, Zheng Shi, and Hong Wang
\thanks{Wenshuai Liu  and Yaru Fu are with the School of Science and Technology, Hong Kong Metropolitan University, Hong Kong 999077, China (e-mail:~liuws1996@gmail.com,~ yfu@hkmu.edu.hk). \emph{(Corresponding author: Yaru Fu)}}
\thanks{Zheng Shi is with the School of Intelligent Systems Science and Engineering, Jinan University, Zhuhai 519070, China (e-mail: zhengshi@jnu.edu.cn).}
\thanks{Hong Wang is with the Key Lab of Broadband Wireless Communication and Sensor Network Technology of Ministry of Education, Nanjing University of Posts and Telecommunications, Nanjing 210003, China (e-mail: wanghong@njupt.edu.cn).} 
}
}

\maketitle

\begin{abstract}
The convergence of digital twin technology and the emerging 6G network presents both challenges and numerous research opportunities. This article explores the potential synergies between digital twin and 6G, highlighting the key challenges and proposing fundamental principles for their integration. We discuss the unique requirements and capabilities of digital twin in the context of 6G networks, such as sustainable deployment, real-time synchronization, seamless migration, predictive analytic, and closed-loop control. Furthermore, we identify research opportunities for leveraging digital twin and artificial intelligence to enhance various aspects of 6G, including network optimization, resource allocation, security, and intelligent service provisioning. This article aims to stimulate further research and innovation at the intersection of digital twin and 6G, paving the way for transformative applications and services in the future.
\end{abstract}


\section{Introduction} \label{sec:introduction}
The comprehensive deployment of 5G has driven exploration into 6G, propelling wireless communication technology from ``Ubiquitous Connectivity" towards ``Intelligent Connectivity". 
This transformation has given rise to numerous emerging applications and services such as holographic communication, multidimensional sensing, and ubiquitous intelligence. 
To meet the demands of these new applications and services, 6G faces tremendous challenges, including ultra-dense network deployment, massive ultra-reliable and low latency communications, and self-sustaining system\cite{Letaief2022JSAC,Deng2023WCNC}.
In light of these challenges, digital twin (DT) is being considered as a powerful candidate for driving the design, analysis, and operation of 6G, thanks to the rapid advancements in artificial intelligence (AI), high-performance computing, and semiconductor fabrication technology. 
By leveraging real-time monitoring and interaction with the physical world, DT enables the creation of highly accurate digital replicas of the physical space.
Furthermore, the integration of AI empowers DT to enhance decision-making processes within the physical networks\cite{Lin2023MCOM, Wang2022IoT_Mobility}.

Wireless digital twin networks (WDTNs) play an essential role as high-fidelity digital replicas of the entire lifecycle of the 6G physical network. 
The proposed architecture integrates seamlessly with ongoing efforts on standardized network architectures.
WDTNs enhance the 6G ecosystem by continuously evaluating the efficiency and effectiveness of quality of service (QoS) strategies via real-time monitoring, simulation, and analysis of the physical network. 
DT-driven approach not only optimizes and predicts the behavior of the physical system to ensure optimal performance, but also supports proactive decision-making. Furthermore, the architecture is designed to facilitate cooperation and coexistence with standardized network frameworks, ensuring it complement and extend existing and emerging standards.
Moreover, WTDNs integrate multi-access edge computing (MEC) for the deployment of DTs, utilizing communication and computational collaboration with multiple edge servers and end devices to effectively reduce communication latency. 
Meanwhile, DT supported by AI facilitates the decision-making process through real-time interaction between the digital and physical realms \cite{Khan2022MCOM}.

Several notable research work have explored the role of DTs in optimizing different aspects of WDTNs. 
For example, the work \cite{Dai2021TII} applied AI techniques into DT to enhance the design of edge computing networks. By mirroring the dynamic network in DT, AI is applied to implement real-time resource allocation, which effectively improves energy efficiency and data processing efficiency.
In \cite{Li2022TVT}, an adaptive aerial edge computing network was proposed. The network utilized DT to achieve real-time environmental predictions, guiding intelligent offloading strategies and resource allocation. As a result, the network's energy consumption was significantly reduced. 
In \cite{Liu2022IOTJ}, the authors integrated DT with blockchain to improve network security, in which an intelligent task offloading algorithm was developed to reduce delay and energy consumption while ensuring data security.
Another hot research issue is to ensure the provision of reliable and accurate WDTNs. This requires careful consideration of real-time synchronization between the physical devices and their corresponding DTs. 
In \cite{Vaezi2023IOTJ}, the authors developed a model for DT synchronization, aiming to minimize the age of information associated with DT data. 
The authors in \cite{Zheng2023TWC} investigated DT synchronization problem in vehicular networks and analyzed the association between DT and wireless access points. The objective was to achieve low-latency synchronization between DTs and access points.
Furthermore, in \cite{Chukhno2022IOTJ}, the authors considered social characteristics, computation resources, and user mobility in their analysis of DT deployment. They proposed an approximation algorithm to minimize the synchronization delay.
Last but not least, the mobility of mobile users (MUs) may lead to scenarios where direct synchronization between physical users and their DTs is not feasible. In such cases, DT migration provides a solution to maintain the DT's functionality and ensure that it remains responsive and up-to-date. Here, DT migration refers to the process of moving a DT from one server to another within the network. The migration process involves transferring the DT's data and reconfiguring it in the new environment.
The authors in \cite{Lu2021IOTJ} studied DT deployment and migration problems using AI techniques. They aimed to maximize MU utility by achieving a balance between latency and energy costs.

While existing studies have proposed various real-time decision-making algorithms, they have primarily focused on the short-term deployment and synchronization of DT, overlooking the long-term evolution requirements of DT.
By integrating DT deployment, synchronization, and migration, the long-term operation of DT can be achieved efficiently. This allows for the comprehensive balancing of multiple objectives, such as resource utilization, QoS, and adaptability, thereby providing a consistent and sustained DT service experience.
Furthermore, these studies have not comprehensively considered the incorporation of advanced AI techniques and the powerful computing and communication capabilities offered by 6G networks.
In this article, we present a comprehensive review of WDTNs from the perspective of 6G, starting with the design of a novel architecture. 
Our proposed architecture integrates  advanced AI technologies, including semantic communications (SC), large language models (LLMs), and deep reinforcement learning (DRL), along with 6G technologies like terahertz (THz) communication and reconfigurable intelligent surfaces (RIS). By incorporating these technologies, precise mapping and real-time synchronization of complex dynamics within WDTNs are achieved, enhancing DTs with autonomous learning and decision-making abilities.
Moreover, the devised architecture empowers WDTNs with self-learning capabilities, enabling comprehensive analysis and tailored strategies for complex network managements and services.
To demonstrate the effectiveness of our proposed approach, we present a case study on reliability-aware synchronization for WDTNs and provide numerical results to validate its performance.

The remaining sections of this work are organized as follows.  
Section \ref{s:framework} provides a meticulously designed architecture for WDTNs and discusses the key issues inherent to it. 
Additionally, potential enablers for addressing these issues are also explored. In Section \ref{s:case}, we present a case study that investigates the synchronization mechanisms within WDTNs, validating the improvements in energy saving and reliability achieved through the proposed AI-driven solution. Lastly, Section \ref{s:future} concludes this article and outlines future research directions in this field.

\begin{figure*}[t]
	\centering
	\includegraphics[width=0.9\textwidth]{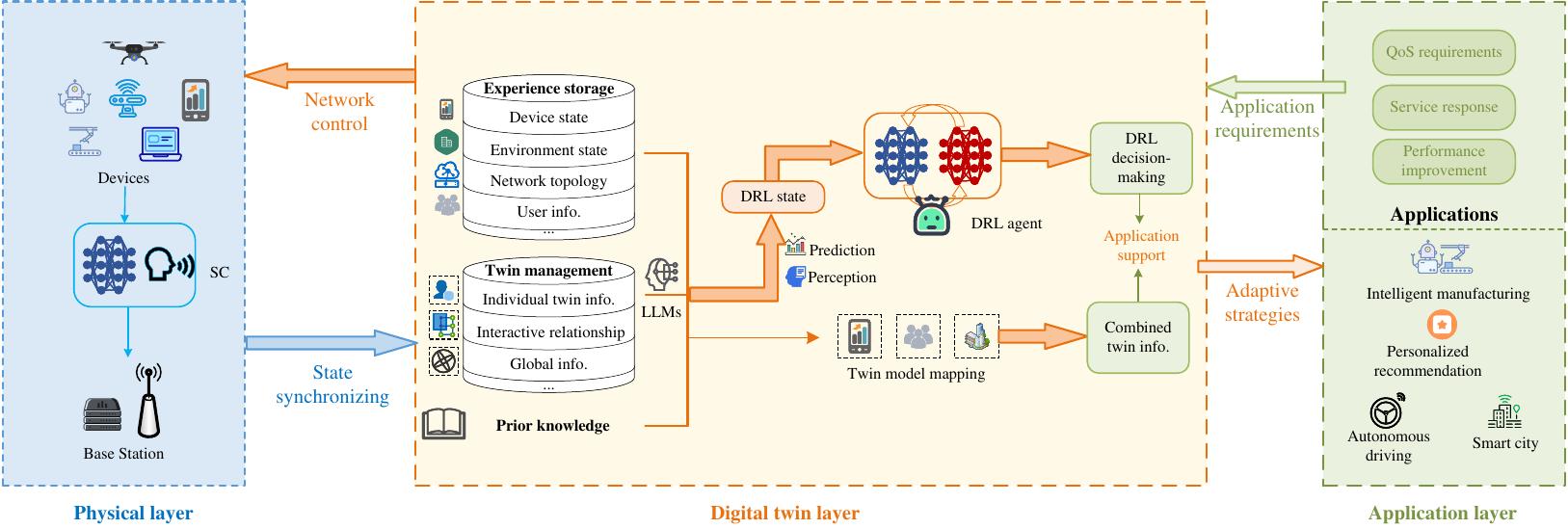}
	%\vspace{-2em}
	\caption{ An illustration of the WDTNs framework.}
	\label{fig:wdframework}
	\vspace{-1.5em}
\end{figure*}
\section{Wireless Digital Twin Networks}\label{s:framework}
In this section, we start by introducing a three-tier network architecture for WDTNs that incorporates various advanced techniques. We then proceed to identify key issues that currently exist within WDTNs.
\subsection{The Architecture of WDTNs}
The WDTNs framework is designed as a three-tier architecture, as depicted in Fig. \ref{fig:wdframework}. This architecture consists of the physical layer, digital twin layer, and application layer. 
The physical layer provides data for the development of the digital twin layer.
With the aid of 6G network, the digital twin layer is responsible for model construction, item mapping, and policy optimization. 
The application layer is a platform designed to deliver comprehensive services and applications. 
The detailed components of the three-tier network are sorted out as follows:
\begin{itemize}
	\item \textbf{Physical layer}: 
The physical layer primarily comprises various devices, base stations (BSs), and other entities that form the communication system. In the context of 6G, advanced supplements such as THz communication and RIS play a significant role in enhancing the capabilities of these entities.
These devices utilize the THz frequency range for high-speed data transmission to gather raw data. By leveraging the beamforming and reflection properties of RIS, these data are efficiently transmitted to the BS and further uploaded to the digital twin layer. The digital twin layer extensively processes the received raw data to create information that accurately maps the state of the environment, facilitating real-time interaction and control with the physical layer. It is noteworthy that BSs are strategically positioned in close proximity to the devices in order to provide wireless interfaces, ensuring seamless and instant communication within the network. These BSs also incorporate edge servers, which greatly enhance the computational and data processing capabilities of the devices.
In addition, SC is used to optimize the collected data, ensuring that only information relevant to specific task objectives is conveyed. This optimization of data transmission reduces communication overhead and enhances the overall operational efficiency of the system.
	\item \textbf{Digital twin layer}:
	The digital twin layer is the core of the WDTNs architecture, serving as a bridge between the physical and the application layers. 
	It ensures the efficient operation of the network by comprising three key elements: data storage, virtual model mapping, and twin management.
	Firstly, data storage collects key data, including real-time environmental inputs, experiential insights, and historical records, providing the foundation for model creation and refinement. 
	Secondly, virtual model mapping precisely replicates the physical network using the filtered data, ensuring that the digital twin reflects its state in real-time and predicts future performance. 
	In this process, LLMs utilize their capabilities to transform complex data into more easily understandable information, helping managers make faster and wiser decisions.
	Lastly, twin management is responsible for supervising the real-time synchronization between the digital twin and its physical counterpart. 
	DRL technology plays a crucial role, achieving automated optimization and maintenance of the network through continuous learning and strategy adjustments.
	The combination of these elements not only makes the digital twin layer become the neural center for managing WDTNs but also endows wireless networks with unprecedented adaptive capabilities and intelligent decision-making. 
	As a result, the network becomes more stable, responsive, and capable of flexibly addressing future technological advancements and the challenges of complex scenarios.
	\item \textbf{Application layer}:
	The application layer integrates information from the physical layer and data from the digital twin layer to support intelligent decision-making and service responsiveness. 
	For instance, in intelligent manufacturing, the application layer can respond in real time to changes in the production line and adjust manufacturing strategies; 
	in personalized recommendation systems, it analyzes user information and behavior data to provide customized content; 
	and in autonomous driving and smart city applications, it processes vast amounts of environmental and interactive data to ensure safety and efficiency. 
\end{itemize}
The interplay among these three layers is vital. 
In particular, the  interaction between the physical layer and the digital twin layer consists of gathering data and managing the network. 
Accurate and complete data from the physical components is crucial, as it underpins the creation of the digital twin layer. 
The digital twin layer, through simulation and optimization, offers efficient and enhanced strategies for the management of physical network. 
Meanwhile, the digital twin layer and the application layer interact to set up applications and cater to individual service demands.

\subsection{Key Issues and Potential Enablers }
In the considered WDTNs paradigm, achieving real-time model synchronization and intelligent mirroring of physical entities is of paramount importance. This requires the digital twin layer to possess self-optimization, self-configuration, and self-healing capabilities to accurately map physical entities to virtual twins. Specifically, to achieve the precise mapping of physical entities to virtual twins in WDTNs, we address the following key issues to enhance their adaptability.
\begin{itemize}
	\item \textbf{SC in WDTNs}: 
	SC meticulously filters the content of data transmission, ensuring that only information substantially contributing to the objectives of a task is conveyed. 
	This strategy significantly alleviates communication load, thereby enhancing the system’s overall efficiency.  
	However, a challenge encountered in this process is that the core of the WDTNs is to construct digital replicas of physical entities in the virtual space. 
	Introducing SC into this requires considering how to map the semantic information of the physical world to the virtual world, and how to feed back the semantic information of the virtual world to the physical world. 
	This necessitates the development of new semantic mapping models and mechanisms to achieve seamless integration between the physical and virtual worlds. 
	Moreover, physical entities and DTs need to be synchronized in real time. SC optimizes state synchronization by filtering and transmitting relevant data and its semantics, reducing transmission load. By employing SC for state synchronization, the transmitted data becomes more meaningful, ensuring improved system efficiency.
    However, this requires the synchronization mechanism to ensure real-time performance, consistency, and integrity of all synchronized information, including semantic data.
	
	\item \textbf{LLMs in WDTNs}: 
	LLMs enhance intelligent interaction in the digital twin layer by understanding and generating text, improving user experience and facilitating seamless human-machine interaction.
        Furthermore, they aid in data analysis and decision-making by extracting insights from extensive textual data. 
  	However, LLMs require tremendous computational power, including high-performance processors and large memory capacities. 
   To address these challenges, techniques such as model compression, model splitting, edge-cloud collaboration and adaptive inference can be employed to enable the deployment of LLMs on resource-constrained edge devices.
	Another crucial issue is the handling of sensitive information and personal data when LLMs process and generate text. 
	To mitigate privacy concerns, techniques such as differential privacy, federated learning, and secure multi-party computation can be applied to ensure that sensitive information is protected during the training and inference of LLMs.
	
	\item \textbf{THz and RIS in WDTNs}: 
	 THz and RIS are introduced as key enabling technologies in the physical layer. 
	THz offer abundant spectrum resources, enabling wireless transmission at unprecedented speeds. 
	However, they face challenges such as high directionality, propagation losses, and blockages, which may lead to signal degradation and service interruptions. 
	To address these issues, high-gain directional antennas, advanced beam tracking and alignment mechanisms, and adaptive modulation and coding schemes should be employed. 
	Furthermore, RIS is leveraged to enhance the propagation and coverage of THz signals by intelligently shaping the wireless environment. 
	However, in the WDTNs, it is necessary to consider how to map the physical channel characteristics of THz and RIS to the virtual world to construct accurate DTs. 
	This requires the development of new physical-virtual channel modeling techniques to capture the key characteristics of THz and RIS channels and transform them into parameters and behaviors in the DT. 
	Moreover, the introduction of THz and RIS makes the wireless channel more dynamic and complex. 
	New real-time channel estimation, tracking, and updating mechanisms need to be developed to ensure the accuracy and timeliness of the DT.

	\item \textbf{Synchronization in WDTNs}:
	Efficient synchronization of DTs aims to ensure precise matching between physical entities and virtual entities in terms of their states and behaviors, thereby reflecting environmental information and enhancing the system's cognition, reasoning, and decision-making capabilities.
	Physical entities collect real-time data when changes occur in themselves or in external environmental conditions.
        Subsequently, these data are transmitted to the digital twin layer.
	The collected data, combined with historical simulation results, experiences, and memories, is used to drive the self-updating and learning process of WDTNs.
    During the synchronization process, leveraging the capabilities of LLMs for semantic parsing and advanced analysis of complex data. 
	Then, DRL can make more precise resource scheduling decisions based on rich contextual, thereby enhancing the flexibility and efficiency of the network.

	\item \textbf{Deployment in WDTNs}:
	The sustainable and enduring deployment of DTs is important and a key driver in advancing WDTNs toward intelligence and automation. 
	It reduces delays and energy consumption during the synchronization process.
	Meanwhile, this strategy optimizes the usage of edge server resources, narrows the gap between physical and digital systems, and enhances the precision of synchronization.
         However, a significant challenge is ensuring the long-term deployment of DTs remains robust and adaptable in the face of evolving environmental conditions.
	By employing DRL and LLM, the system dynamically adjusts deployment strategies based on network load and DT characteristics. 
 
	\item \textbf{Migration in WDTNs}: 
    DT migration can customize the user experience by transferring the DT to nearby edge computing nodes based on the devices' actual location and personal preferences, providing a faster and tailored user experience. 
	With continuous changes in the environment and user demands, DT migration allows the system to adaptively adjust, ensuring both efficient synchronization and accuracy of the DT. 
	In addition, DT migration enhances system reliability and robustness by migrating to other nodes in the event of hardware failures or node overloads, thus preventing service disruptions. 
	Unlike the initial deployment of the DT, migration strategies focus more on continuously optimizing the deployment scheme over time. 
        However, DT migration poses several challenges. Ensuring seamless migration without service interruptions demands intricate coordination among various network nodes, which can heavily utilize resources. Another challenge is the need for real-time decision-making to determine the optimal time and deployment for migration.
        To address these challenges, the system utilizes DRL for decision-making to determine the optimal times and conditions for DT migration, considering various network states and operational metrics. Moreover, LLMs are employed to analyze patterns and trends from vast amounts of operational data. This enables the prediction of potential node failures or overloads, allowing the system to initiate preemptive migrations and enhance reliability while reducing the risk of service disruptions.
\end{itemize}

\begin{figure*}[t]
        \centering
	\includegraphics[width=0.75\textwidth]{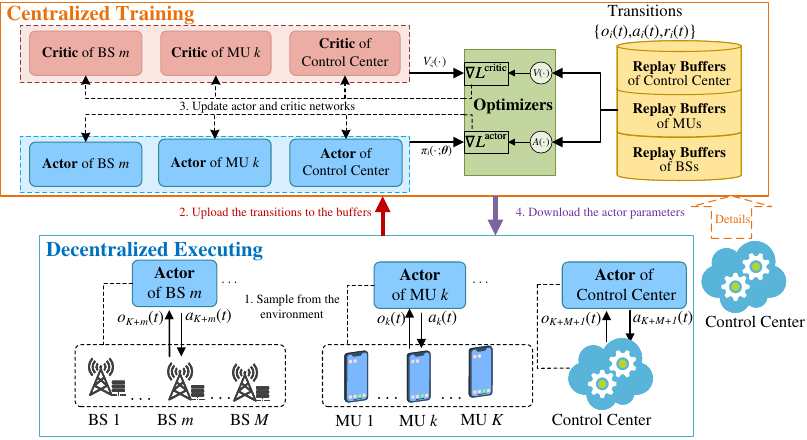}
	\caption{An illustration of the DRL-based training framework.}
	\label{fig:framework}
	\vspace{-1.5em}
\end{figure*}
\section{A Case Study}\label{s:case}
The previous discussions have demonstrated the importance of employing machine learning solutions to address the key issues in WDTNs. In this section, we present a case study of 6G that specifically focuses on DT synchronization within WDTNs, considering the challenges posed by the highly dynamic nature and stringent reliability requirements in 6G networks. This case study highlights how machine learning-driven solvers can effectively tackle the synchronization problem, providing a practical solution for achieving efficient DT synchronization in WDTNs.

\subsection{System Model} 
We consider a WDTN consisting of a control center, multiple BSs, and MUs.
The control center is used to maintain the model training and the deployment decision of DTs.
The movement of MUs follows a Gauss-Markov process, while the DT synchronization requests of MUs follow a Bernoulli distribution. Moreover, the wireless channels between MUs and BSs are modeled as Rician fading channels.
Each BS is equipped with a server to assist MUs in constructing and synchronizing their DTs.
To ensure the fidelity of WDTNs, the servers in BSs require synchronization of information from the MUs. 
Furthermore, the MUs may switch their BS associations after moving for a period of time which makes DT migration become imperative.
However, the occurrence of synchronization failure cannot be completely avoided.
Specifically, when a MU initiates a synchronization request, it may encounter failure under two conditions. Firstly, if the MU's DT is undergoing migration, it may be unable to establish synchronization successfully. Secondly, if the required synchronization time exceeds the latency requirement, the synchronization request is aborted. In this study, our objective is to minimize the long-term average energy consumption of the MUs taking into account the following three aspects: 1) Ensuring the reliability of DT synchronization.
2) Optimizing the deployment of DT, and 3) Efficiently allocating computational and communication resources.
Considering the inherent structural properties of the optimization problem, we 
propose a learning-based solution. Specifically, we define the agent, state, action, and reward as follows:
\begin{itemize}
    \item \textbf{Agent}:	it includes three types of agents, i.e., a central controller agent, multiple BS agents and MU agents.
    \item \textbf{State}:	the state includes the positions of MUs, the locations of BSs, and the computational resources of edge servers.
    \item \textbf{Action}: we design distinct actions for the three types of agents. MU agents decide on their transmission power. BS agents optimize the computational resource allocation. Whilst, the central controller decides on the deployment of DT.
\item  \textbf{Reward:} the reward is defined as the weighted sum of energy consumption and reliability. We use $\eta$ to control the weights of energy consumption and reliability.
\end{itemize}
With foregoing definitions, we elaborate on our proposed training framework, which is based on a heterogeneous agent proximal policy optimization with Beta distribution (Beta-HAPPO)  \cite{twc_liu}, where each MU, BS, and the control center are represented as individual agents. Each agent possesses an actor network and a critic network, as depicted in Fig. \ref{fig:framework}.
The HAPPO algorithm is a policy-based trust region learning scheme that utilizes a centralized training and decentralized execution mechanism, enabling more effective collaborative learning. Unlike other approaches, HAPPO does not require agents of the same type to share identical neural network parameters with their distributed actor and critic networks.
Thus, it prevents the exponential degradation of the reward function that usually happens as the number of agents increases.

The detailed training processes consist of two main phases: environmental interaction and policy update. In the environmental interaction phase, the agent collects the current network status and transmits this state information to the actor and critic networks. The actor network makes actions based on this information. Thereafter, the environment provides rewards based on the agents' actions and stores this data in the replay buffer. 
During the policy update phase, a batch of data samples is extracted from the experience replay buffer to train the actor and critic networks. Subsequently, the critic network evaluates the value function, aiding the actor network in optimizing its policy. Through this process, the agents continuously learn and adjust their policies for better interaction with the environment, enhancing the system's performance.

\subsection{Numerical Results}
We evaluate the performance of our proposed Beta-HAPPO scheme by numerical simulations.
The parameter settings are presented as follows. We consider a 1000 m $\times$ 1000 m square area, where the MUs are uniformly distributed. We set the number of MUs as 30 and the number of BSs as 5. 
Other default settings for environment and algorithm are typically based on the parameters in\cite{Liang2022TWC, Waqar2022TITS}. 
For performance comparison, we focus on DRL approaches, as they provide a robust framework for dynamically adapting to complex network environments. The following benchmark schemes are used:
\begin{itemize}
	\item \textbf{Gaussian-HAPPO:} HAPPO adopts Gaussian distribution in actor networks.
	\item \textbf{Beta-MAPPO:} Multi-agent proximal policy optimization (MAPPO) adopts Beta distribution in actor networks. Be noted that unlike HAPPO, agents need to share parameters among themselves.
	\item \textbf{MADDPG:} Multi-agent deep deterministic policy gradient (MADDPG) employs a deterministic policy with exploration noise and does not utilize distribution as the actor's output.
\end{itemize}

\begin{figure}[t]
	\centering
	\subfigure[{Average energy consumption vs. time steps.}]{
		\begin{minipage}[t]{0.21\textwidth}
			\centering
			\includegraphics[width=\textwidth]{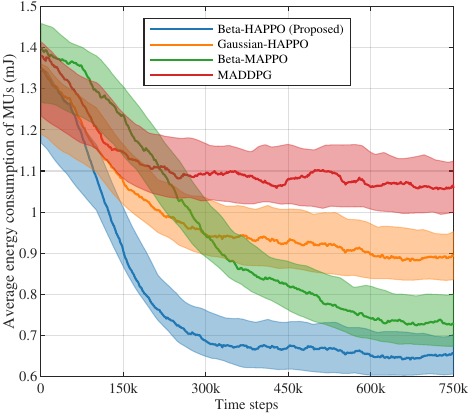} 
		\end{minipage}%
	}\hspace{3mm}% 
	\subfigure[{Average energy consumption and synchronization failure ratio vs. control factor.}]{
		\begin{minipage}[t]{0.23\textwidth}
			\centering
			\includegraphics[width=\textwidth]{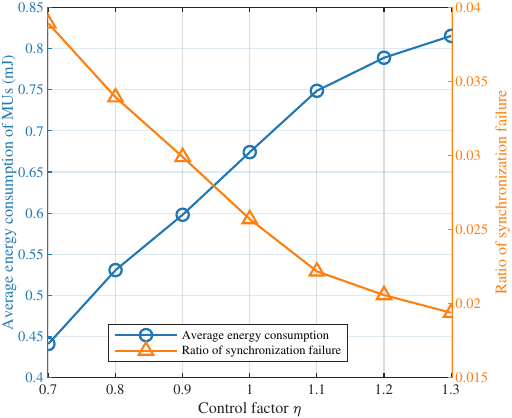}
		\end{minipage}%
	}%
	
	\subfigure[{Average energy consumption vs. number of MUs.}]{
		\begin{minipage}[t]{0.21\textwidth}
			\centering
			\includegraphics[width=\textwidth]{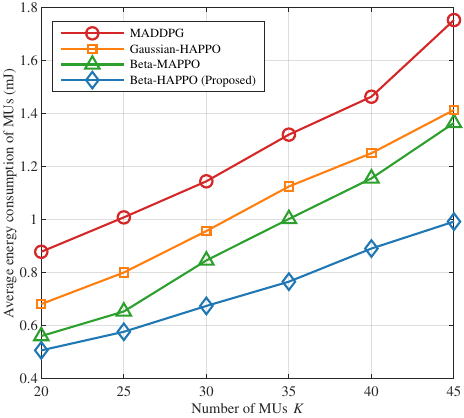}
		\end{minipage}
	}\hspace{3mm}%
	\subfigure[{Synchronization failure ratio vs. number of MUs.}]{
		\begin{minipage}[t]{0.21\textwidth}
			\centering
			\includegraphics[width=\textwidth]{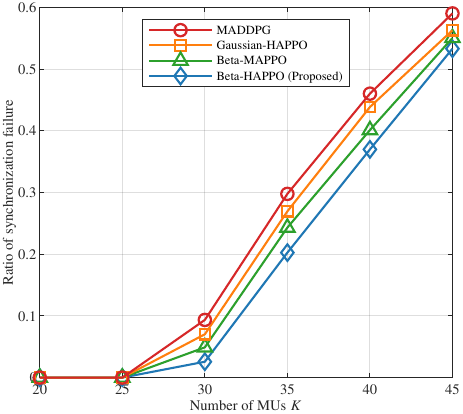}
		\end{minipage}
	}%
	\centering
	\caption{Energy consumption and synchronization failure ratio trends.}
	%\vspace{-1em}
\end{figure}

Fig. 3(a) shows the average energy consumption versus time steps. 
It can be observed from Fig. 3(a) that the average energy consumption steadily decreases as the time steps increase, indicating the effectiveness of DRL schemes. 
Moreover, Beta-HAPPO is better than Beta-MAPPO because HAPPO does not employ a parameter sharing strategy.
Furthermore,  HAPPO-based schemes exhibit significantly faster convergence speed compared to the MADDPG scheme. %the convergence speed of HAPPO-based schemes significantly outperform the MADDPG scheme. 
This is because the deterministic policy of MADDPG has insufficient exploration, resulting in insufficient quality and limited diversity of samples in the replay buffer, which hinders the learning progress and slows down the convergence speed.
Moreover, Gaussian-HAPPO requires a higher energy consumption compared to Beta-HAPPO, validating that the Beta distribution is better than Gaussian distribution in the scenarios with bounded actions.
In Fig. 3(b), we evaluate the impact of control factor $\eta$ on the energy consumption and synchronization failure ratio of MUs. As $\eta$ increases from 0.7 to 1.3, we observe that the energy consumption of MUs increases while the synchronization failure ratio decreases. This demonstrates the feasibility of incorporating the control factor to balance the trade-off between energy consumption and synchronization failure of MUs.
Fig. 3(c) shows the average energy consumption versus the number of MUs.
It can be observed that the average energy consumption of MUs gradually increases as the number of MUs increases.
As the number of MUs increases, signal interference among them rises, consequently requiring higher transmission power and resulting in increased average energy consumption.
Fig. 3(d) shows the synchronization failure ratio versus number of MUs.
It can be observed that when the number of MUs is within 25, the synchronization failure ratio is 0; when the number of MUs exceeds 30, the synchronization failure ratio increases as the number of MUs increases. 
When the number of MUs is small, the network resources are sufficient to meet their DT synchronization needs. However, as the number of MUs increases without a corresponding increase in network resources, resources become exhausted, leading to insufficient allocation for DT synchronization and a rise in the failure ratio.

\section{Conclusion and Research Prospects}\label{s:future}
In this article, we conducted a comprehensive review for the architecture of WDTNs, exploring the integration of DT technology with future generation communication systems. We explicitly identified and discussed the intrinsic key issues associated with WDTNs, along with promising solutions. While the construction of WDTNs facilitates the development of data-driven and self-learning networks, leading to increased intelligence and automation, several challenges hinder their real implementations in practice. The complexity of network environments, uncertainty of data, and resource limitations in 6G wireless networks pose significant obstacles. Addressing these challenges is essential for the successful implementation of WDTNs. 
Here are some key challenges that require future attention in terms of efficiency, reliability, and real-world application. The efficiency aspect encompasses considerations related to data processing, resource utilization, and energy efficiency:
\begin{itemize}
\item {Data preprocessing:} 
{Data preprocessing is a critical step in constructing DTs models, providing a foundation of data for accurate DT synchronization.
However, raw data often contains corrupted, improperly formatted, duplicated, or missing information. 
Consequently, thorough preprocessing and cleansing of the data are necessary to ensure data cleanliness and availability, thereby facilitating the precise construction and synchronization of DTs.}
%\textbf

\item {Resource management:}
{Resource management in WDTNs is complex and challenging.  This is due to the intricacy and dynamism of the radio environments in 6G communication networks. Moreover, the access of a large number of users further increases the burden on network resources. This necessitates the effective integration of machine learning algorithms into resource management to achieve adaptive and efficient network operations. }

\item {Energy consumption:}
{The energy consumption issue in WDTNs is an urgent problem that needs to be resolved. This is due to the high computational demands of AI technologies. Moreover, the energy-intensive nature of dense network significantly exaggerates this challenge.
To address this, measures such as implementing novel energy management and scheduling schemes, adopting green energy sources, applying power consumption optimization techniques, and hardware designs can be considered.}
\item {Fault-tolerant system:}
{A fault-tolerant system is important for WDTNs. Due to the transmission of erroneous data between DT entities, the integrity of the entire digital twin layer could be compromised, thereby reducing the performance of the network. Therefore, effective fault tolerance mechanisms must be designed to mitigate the impact of faults and shorten the recovery times.}
\item {Pilot studies or real-world trials:}
 {Conducting pilot trials and real-world experiments is crucial.
Deploying the algorithm in real WDTNs will provide valuable insights into its performance under practical constraints and dynamic conditions.
These trials will identify potential challenges and limitations that may not be apparent in simulations, such as hardware imperfections, communication delays, and scalability issues.
Real-world validation will enable fine-tuning of the algorithms and adaptation to specific application scenarios.
In order to bridge the gap between simulation and reality, pilot trials will pave the way for successful implementation of the proposed approach in practical WDTNs. }
\end{itemize}

\bibliographystyle{IEEEtran}
\bibliography{IEEEabrv,refs}

\end{document}